\documentclass[superscriptaddress,twocolumn,nofootinbib,showpacs,prl,amsmath]{revtex4}
\usepackage{times}
\usepackage{graphicx}
\usepackage{longtable}
\usepackage{color}
\usepackage{amsmath}
\usepackage{amssymb,srcltx}

\newcommand{\beq}{\begin{equation}}
\newcommand{\eeq}{\end{equation}}
\newcommand{\beqa}{\begin{eqnarray}}
\newcommand{\eeqa}{\end{eqnarray}}
\newcommand{\ket} [1] {\vert#1\rangle}

\def\ket#1{|#1\rangle}

\def\half{\frac{1}{2}}

\newcommand{\HE}{\ket{{\rm HE}_6}}
\newcommand{\CZ}{\textsf{CZ}}

\newcommand{\HH}{\textsf{H}}

\newcommand{\HEtilde}{\ket{\widetilde{\rm HE}_6}}
\newcommand{\E}{\ket{{\textsf{E}}}}
\newcommand{\Etilde}{\ket{\widetilde{\textsf{E}}}}

\begin{document}

\title{Experimental Realization of the Deutsch-Jozsa Algorithm with a Six-Qubit Cluster State}
\author{Giuseppe Vallone}
\homepage{http://quantumoptics.phys.uniroma1.it/}
\affiliation{Museo Storico della Fisica e Centro Studi e Ricerche Enrico Fermi, Via Panisperna 89/A, Compendio del Viminale, 00184 Roma, Italy}
\affiliation{Dipartimento di Fisica, Universit\`{a} Sapienza di Roma, 00185 Roma, Italy}
\author{Gaia Donati}
\homepage{http://quantumoptics.phys.uniroma1.it/}
\affiliation{Dipartimento di Fisica, Universit\`{a} Sapienza di Roma, 00185 Roma, Italy}
\author{Natalia Bruno}
\homepage{http://quantumoptics.phys.uniroma1.it/}
\affiliation{Dipartimento di Fisica, Universit\`{a} Sapienza di Roma, 00185 Roma, Italy}
\author{Andrea Chiuri}
\homepage{http://quantumoptics.phys.uniroma1.it/}
\affiliation{Dipartimento di Fisica, Universit\`{a} Sapienza di Roma, 00185 Roma, Italy}
\author{Paolo Mataloni}
\homepage{http://quantumoptics.phys.uniroma1.it/}
\affiliation{Dipartimento di Fisica, Universit\`{a} Sapienza di Roma, 00185 Roma, Italy}
\affiliation{Istituto Nazionale di Ottica
Applicata (INOA-CNR), L.go E. Fermi 6, 50125 Florence, Italy}

\date{\today}


\begin{abstract}
We describe the first experimental realization of the
Deutsch-Jozsa quantum algorithm to evaluate the properties of a
2-bit boolean function in the framework of one-way quantum
computation. For this purpose {a novel
two-photon six-qubit cluster state was engineered. Its peculiar
topological structure is the basis of the original
measurement pattern allowing the algorithm realization.} The good
agreement of the experimental results with the theoretical
predictions, obtained at $\sim$1kHz success rate, demonstrate the
correct implementation of the algorithm.
\end{abstract}



\maketitle


{\em Introduction. --} In the last decade, quantum information
processing and, in particular,  quantum computation, have been conquering increasing
interest and importance in the scientific community, supported by
the promising theoretical and experimental results obtained. One
of the many present efforts is the construction of quantum
hardware, which up to now has been realized by following different
experimental techniques \cite{kok07rmp,benh08nap}. In this way, it was then possible to
demonstrate the correct functioning of one and two-qubit logic
gates as well as the successful implementation of quantum
algorithms which strongly show the efficiency of a quantum
computer with respect to its classical analogue. Among these, the
Deutsch-Jozsa (DJ) algorithm is the first example of the speed-up
exhibited by a computer taking advantage of quantum mechanics in
the evaluation of a global property of an $n$-bit boolean
function \cite{deut02pro}.

In this Letter we report the realization of the Deutsch-Jozsa
algorithm in the framework of the one-way model of quantum
computation \cite{raus01prl,brie09nap}, which has
already proved successful in the
construction of quantum gates such as the controlled-\textsc{NOT}
(\textsc{CNOT}) gate \cite{vall08prl,gao09qph, vall09qph} and in
the implementation of the Grover \cite{vall08pra,
walt05nat,prev07nat, vall08prl} and the Deutsch algorithms
\cite{vall08pra,tame07prl}. The latter corresponds
to the case $n=1$ and is based on the use of
four-qubit cluster photon states. Here we get the access to the
case $n = 2$ by taking advantage of a peculiar two-dimensional
two-photon six-qubit cluster state generated by a source of
multi-qubit cluster states whose performances have been already
demonstrated
\cite{cecc09prl,vall09qph}. At variance with the simple case $n =
1$, the DJ algorithm allows to take advantage of the exponential
growing of the computational speed-up for
increasing values of  $n$, as said. Hence the results presented in
this paper are important in that they open the way to the
implementation of the DJ algorithm with a still larger number of
qubits. Although the DJ algorithm has been implemented before with
photons \cite{brai03prl}, our realization represents the first
realization with a 2-bit function in the context of
{measurement-based quantum computation}.

\begin{figure}[t]
\begin{center}
\includegraphics[width=8.5cm]{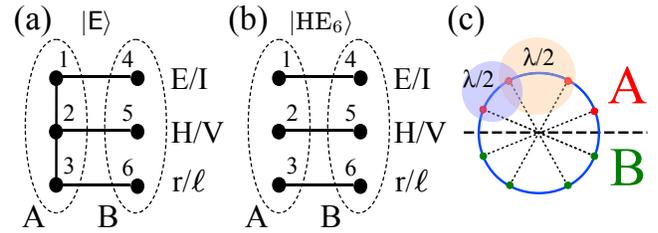}
\caption{Graphs associtated to (a) the $\E$ cluster and (b) the
$\HE$ hyperentangled state. Qubits 1, 4 are encoded in the E/I
momentum, qubits 2, 5 are encoded in polarization (H/V) and qubits 3, 6 
in the r/$\ell$ momentum. (c)
Annular section of the conical SPDC emission of a Type I phase
matched crystal. The source produces the $\HE$ state over eight
spatial modes. Two half wave-plates ($\lambda/2$) with vertical
optical axis intercepting three modes of the $A$ photon are used
to transform $\HE$ into the $\E$
state.} \label{fig:graphs}
\end{center}
\end{figure}
{\em Realization pattern for the Deutsch-Jozsa algorithm. --} Let
us briefly recall the generalized version of the Deutsch
algorithm \cite{deut02pro}, where a set of $n$ qubits constitutes the input of a
black box, usually known as the Oracle, which implements the
$n$-bit boolean function $f(x)$ such that $f: \{0,1\}^n
\rightarrow \{0,1\}$. The aim of the DJ algorithm is to
determine whether the function evaluated by the oracle is constant
or balanced; a function $f$ is said to be balanced if it is equal
to $0$ when calculated in half of the possible values of $x$ and
equal to $1$ when the remaining allowed values for $x$ are taken
into account. Classically, $2^{n-1} + 1$ queries to the oracle are
necessary to solve the problem while, in the frame of quantum mechanics,
the answer comes with one single query. Consequently, the greater
is the number $n$ of qubits involved, the more evident is the
difference in the performances of the quantum computer with
respect to its classical counterpart. The initial state of the
system is $\ket{0} \otimes \ket{0} \otimes \cdots \otimes \ket{0}
= \ket{0}^{\otimes n}$ and an ancillary qubit in the state $\ket{1}$
is added to the $n$ input qubits. The operation performed by the
oracle is given by $(\HH^{\otimes n} \otimes \HH)U_f(\HH^{\otimes
n} \otimes \HH)$, where $\HH$ is the Hadamard gate; the unitary operator $U_f$ acts on the states of
the computational basis so that $U_f \ket{x}\ket{y} =
\ket{x}\ket{y \oplus f(x)}$. The final state is found to be
$\bigl(\frac{1}{2^n} \sum_{x, y = 0}^{2^n - 1} (-1)^{f(x)+ x \cdot
y} \ket{y}\bigr) \ket{1}$. Measuring the state of the $n$ qubits in the
computational basis leads to the conclusion: if we get the state
$\ket{0}^{\otimes n}$ the function $f$ is constant, otherwise it
is balanced, as seen from the above expression for the final state
of the global system. Moreover, the measurement of the ancillary
qubit in the computational basis is expected to always return the
$\ket{1}$ state.

We now go into the details of the proposed experimental
realization of the DJ algorithm for { a function acting
on $n = 2$ bits}. In this case, the boolean
function which we are interested in is such that
$f:~\{0,1\}^2\rightarrow\{0,1\}$. The function $f$ can be
calculated in its four arguments $x = 0, 1, 2, 3$, with $x = 2x_1
+ x_0$ and $x_0, x_1 = 0, 1$. Among the 16 possible functions of
this kind, we focus our attention on the balanced function $f_B$,
such that $f_B(0) = f_B(3) = 0$, $f_B(1) = f_B(2) = 1$, and on the
constant function $f_C$ for which we have that $f_C(x) = 0$ for
every allowed value of $x$. We thus identify the state $\ket{x}_Q
= \ket{x_1}_{\overline{1}} \ket{x_0}_{\overline{2}}$ as the input
entering the oracle. In the former expression the subscripts
$\overline{1}$ and $\overline{2}$ refer to logical qubits. As we
know, the implementation of the DJ algorithm requires an
additional ancillary qubit in the initial state $\ket{y}_A \equiv
\ket{y}_{\overline{3}}$\,, where $\overline{3}$ is the logical
qubit associated to the ancilla. For the previously defined
functions we have that $U_{f_C}=\openone$ and
$U_{f_B}=\textsc{Cnot}_{\bar1\bar3} \textsc{Cnot}_{\bar2\bar3}$,
respectively\footnote{$\textsc{Cnot}_{\bar i\bar j}$ indicates a
controlled-NOT gate between logical qubits\ $\bar i$ and $\bar
j$.}.

\begin{figure}[t]
\begin{center}
\centering\includegraphics[width=7cm]{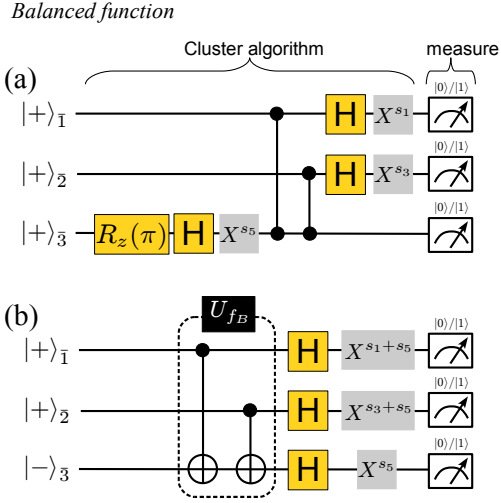}
\caption{Balanced function: (a) algorithm realized
according to the pattern of single-qubit
measurements and (b) equivalent
circuit implementing the DJ algorithm for $n=2$. Gray gates
represent Pauli errors. } \label{fig:circuitB}
\end{center}
\end{figure}
\begin{figure}[t]
\begin{center}
\centering\includegraphics[width=7cm]{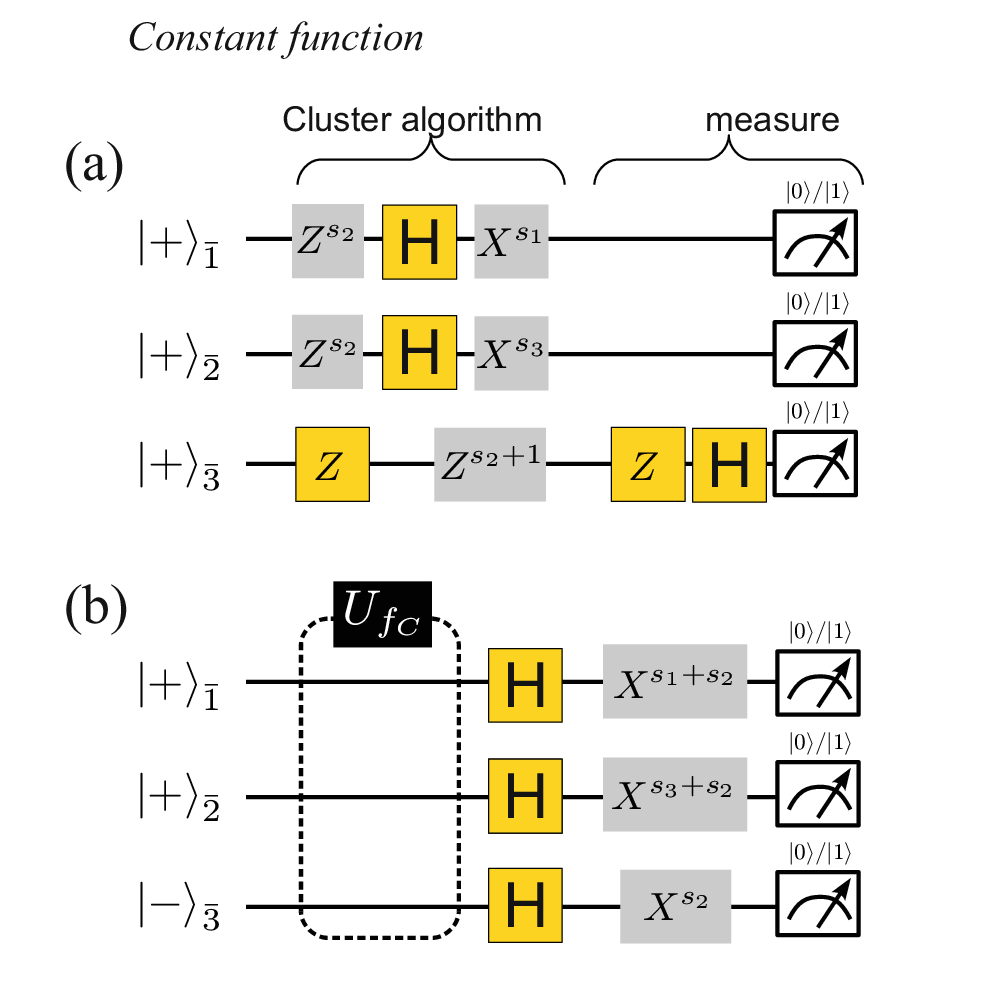}
\caption{Constant function: (a) algorithm realized
according to the pattern of single-qubit
measurements and (b) equivalent
circuit implementing the DJ algorithm for $n=2$. Gray gates
represent Pauli errors. } \label{fig:circuitC}
\end{center}
\end{figure}

In the framework of one-way quantum computing, the starting point
of any computation is the construction of a multi-qubit cluster
state; successively, the choice of a sequence of single-qubit
measurements determines the program to be executed on the quantum
computer. For a review on graph and cluster states and their use
for one-way computing see \cite{raus01prl,brie09nap, hein06var, vall08pra}.
Let us start from the identification of the appropriate
cluster state allowing the realization of the DJ algorithm in the present
work: Fig. \ref{fig:graphs}(a) shows the graph corresponding
to a two-dimensional six-qubit cluster state where the numbered
vertices stand for physical qubits. These qubits are equally
distributed among two photons, labeled as $A$ and $B$:
qubits $1$, $2$ and $3$ belong to photon $A$ and interact by
two controlled-$\textsf{Z}$ gates represented by vertical
connections on the graph, while qubits $4$, $5$ and $6$ are
associated to photon $B$. As usual in the one-way model,
it can be useful to think of the distinct horizontal
qubits as ``the original [logical] qubit at different times'' \cite{niel04prl};
indeed, we identify the
logical qubits $\overline{1}$ and $\overline{2}$ with physical qubits $1$,
$4$ and $3$, $6$, respectively. The ancillary qubit $\overline{3}$
is represented by qubits $2$ and $5$. The ``\textsf{E} cluster''
just described is our quantum computer; we can show that the
choice of the measurement sequence for the two qubits associated
to the ancilla leads to the evaluation of both the balanced $f_B$ and
the constant $f_C$ functions. This implies that, in the \textsf{E}
cluster, qubits $2$ and $5$ play the role of the oracle, while the
remaining qubits constitute the tools we have at our disposal to
discriminate between a balanced and a constant function. To better
understand this feature it is useful to consider the circuit
representations associated to the realization of the two
considered global properties of the 2-bit boolean function $f$.

The proposed measurement configurations are the following:

1. \textit{balanced function -} By measuring qubits $1$, $3$ and $5$ in the
bases $B_1(0)$, $B_3(0)$ and $B_5(\pi)$ we implement, at the logical level, 
the two \textsc{Cnot} gates ($U_{f_B}$) needed to implement the oracle function $f_B$ (see Fig. \ref{fig:circuitB}).
Then we proceed
with the measurement of the output qubits $4$, $6$ and $2$ in the
bases $C_{4}^{\ket{0}}$, $C_{6}^{\ket{0}}$ and
$C_{2}^{\ket{0}}$.

2. \textit{constant function -} We measure qubits $1$, $3$ and $2$ in the
bases $B_1(0)$, $B_3(0)$ and $C_{2}^{\ket{0}}$.
These operations implement, at the logical level, the identity gate $U_{f_C}$ needed to implement the oracle function $f_C$ (see Fig. \ref{fig:circuitC}). Then we read
the result of the computation on the output qubits $4$, $6$ and
$5$ by measuring them in the bases $C_{4}^{\ket{0}}$,
$C_{6}^{\ket{0}}$ and $B_5(\pi)$.

We define $B_j(\alpha) =
\{\ket{\alpha_{+}}_j,\ket{\alpha_{-}}_j\}$ with
$\ket{\alpha_{\pm}}_j = \frac{1}{\sqrt{2}} (\ket{0} \pm
e^{-i\alpha} \ket{1}_j)$, while $C_{j}^{\ket{0}} =
\{\ket{0}_j,\ket{1}_j\}$ is the computational basis for the
Hilbert space associated to qubit $j$. The above sequences of
single-qubit measurement lead us to the circuits shown in Fig.
\ref{fig:circuitB} and \ref{fig:circuitC}; in particular, we can
see the elements of the circuit realizing the unitary
transformation $U_f$ for the balanced function $f_B$ and the
constant function $f_C$, as well as single-qubit Pauli gates. Here
and in the following we indicate with $Z$ ($X$) the Pauli matrix
$\sigma_z$ ($\sigma_x$). For a given basis $B_j(\alpha)$, we
introduce the quantity $s_j$ whose value is $0$ ($1$) if the
measurement result is equal to $\ket{\alpha_{+}}_j$
($\ket{\alpha_{-}}_j$) and equivalently for the $C_{j}^{\ket{0}}$
basis. According to the algorithm, we will expect, as outputs, the
state $\ket{1\oplus s_1\oplus s_5}_{\bar1} \ket{1\oplus s_3\oplus
s_5}_{\bar2}\ket{1\oplus s_5}_{\bar3}$ for the balanced and
$\ket{s_1\oplus s_2}_{\bar1}\ket{s_2\oplus
s_3}_{\bar2}\ket{1\oplus s_2}_{\bar3}$ for the constant function.
In the previous expressions we take into account
the feed-forward corrections of the Pauli errors.

{\em Experimental preparation of the cluster state. --}
Referring to Fig. \ref{fig:graphs}, the two-dimensional cluster
state $\E$ is obtained from a six-qubit hyperentangled state
\cite{vall09pra}, $\HE$, whose graph is shown in Fig.
\ref{fig:graphs}(b) \cite{cecc09prl,vall09pra}. Our experimental
setup adopts a source of two-photon states based on a Spontaneous
Parametric Down-Conversion (SPDC) process where the two particles
are entangled at the same time in the polarization and in two
linear momentum degrees of freedom (DOFs). By a
proper interferometric setup \cite{vall09qph} it is possible to
measure the two spatial DOFs; these variables, labeled as the
``right/left'' momentum ($r/\ell$) and the ``external/internal''
momentum ($E/I$), are both associated to each of the eight modes
on which the two photons are emitted. A detailed description of
the experimental setup, enabling the transformation from the
six-qubit hyperentangled state $\HE$ into a linear cluster state,
can be found in recent papers \cite{cecc09prl, vall09qph}. It is
interesting to note that the transition from the one-dimensional
linear cluster state to the \textsf{E} cluster considered here is
entirely determined by the choice of the controlled-$\sigma_z$
(\CZ) operations corresponding to the vertical links in the graph
(see Fig. \ref{fig:graphs}(a)). These gates are
optically implemented between couples of qubits belonging to the
same photon. The graph associated to the six-qubit cluster state
$\E$ exhibits two links between qubits $1$ and $2$ (encoded in the
$E/I$ momentum DOF and in polarization, respectively) and between
qubits $2$ and $3$ (with qubit 3 encoded in the $r/\ell$ momentum
DOF), hence the corresponding $\CZ_{12}$ and $\CZ_{23}$ logic
gates only involve qubits belonging to photon $A$, as already
noticed above. The optical implementation of the two
controlled-$\textsf{Z}$ gates is realized by means of two
half-wave plates, as shown in Fig.
\ref{fig:graphs}(c).

In order to give an explicit expression for the six-qubit cluster state produced in the present experiment
we point out that the experimental hyperentangled state, which we label as $\HEtilde$,
does not coincide with the hyperentangled state $\HE$ corresponding to the graph in Fig. \ref{fig:graphs}(a)
and instead satisfies the relation
\begin{equation}\label{hyperent}
\begin{split}
\HEtilde &= \textsf{H}_4  Z_5\textsf{H}_5 X_6\textsf{H}_6 \HE =
\\ &= \frac{1}{\sqrt{2}}(\ket{00}_{14} + \ket{11}_{14}) \otimes
\frac{1}{\sqrt{2}}(\ket{00}_{25} - \ket{11}_{25}) \otimes
\\ &\quad \otimes \frac{1}{\sqrt{2}}(\ket{01}_{36} + \ket{10}_{36}),
\end{split}
\end{equation}
where $\textsf{H}_j$ is the Hadamard gate on qubit $j$ and $X_j$ ($Z_j$) is the $\sigma_x$ ($\sigma_z$) gate on the corresponding qubit.
For the \textsf{E} cluster represented by the graph in Fig.
\ref{fig:graphs}(a) we can write that
\begin{equation}\label{Ecluster}
\E = \CZ_{12} \CZ_{23} \HE.
\end{equation}
Combining Eq. \eqref{Ecluster} with Eq. \eqref{hyperent} we get
\begin{equation}\label{Eexp}
\begin{split}
\Etilde &= \CZ_{12} \CZ_{23} \HEtilde=\textsf{H}_4  Z_5\textsf{H}_5 X_6\textsf{H}_6  \E = \\ &=
\half (\ket{EE}\ket{\phi^{-}}_{\pi}\ket{r \ell} +
\ket{EE}\ket{\phi^{+}}_{\pi}\ket{\ell r} \: + \\ &\quad +
\ket{II}\ket{\phi^{+}}_{\pi}\ket{r \ell} +
\ket{II}\ket{\phi^{-}}_{\pi}\ket{\ell r})
\end{split}
\end{equation}
for the six-qubit two-photon \textsf{E} cluster state generated in
the laboratory.  In the above expression the states
$\ket{\phi^{+}}_{\pi}$ and $\ket{\phi^{-}}_{\pi}$ are the two
polarization Bell states. As usual
\cite{vall08prl,walt05nat,prev07nat}, we refer to $\E$ and
$\Etilde$ as the state in the ``cluster'' and ``laboratory''
basis, respectively. As shown in Fig.
\ref{fig:graphs}(c) we transform the hyperentangled state
$\HEtilde$ into the cluster state $\Etilde$ by applying two \CZ\
operations.

{\em Experimental results. --} In order to characterize the
generated $\Etilde$ state we measured the witness operator
$\mathcal W=3 - 2(\prod^3_{k=1}\frac{\widetilde g_{2k} + 1}{2}+
\prod^3_{k=1}\frac{\widetilde g_{2k-1} + 1}{2})$ \cite{toth05pra}
(see \cite{cecc09prl} for the  definition of the $\widetilde
g_i$). We found $\langle \mathcal
W\rangle=-0.333\pm0.002$, demonstrating a genuine
six-qubit entanglement \cite{toth05prl}. Since it is possible to
show \cite{toth05pra, vall07prl} that the fidelity $F$
satisfies the relation $F\geq \frac12(1-\langle
\mathcal W\rangle)$, a lower bound for the fidelity
is easily found: \beq F\geq0.667\pm0.001. \eeq

Let's now turn to the DJ algorithm. We performed the sets of single-qubit measurements stated above
and found the results presented in Table \ref{table:results}: here we show the probabilities of the outputs of the computation
{when no Pauli errors are present (No-FF).
This corresponds to consider only the case
where $s_1=s_3=s_5=0$ for the balanced function and
$s_1=s_2=s_3=0$ for the constant function. We also show the results obtained
by considering all possible outputs and applying} the feed-forward (FF)
operations correcting the Pauli errors (see also Figs. \ref{fig:circuitB} and \ref{fig:circuitC}).
{It is worth noting that, since the output of the computation is read in the $\{\ket0,\ket1\}$ basis, the FF is a
{\it relabeling feed-forward}, i.e. ``the earlier measurement determines the meaning of the final readout''
(see ``Grover's search algorithm'' section of \cite{walt05nat} or the end of section II in \cite{vall08pra}).}

It is also important to notice that the physical
qubits constituting the $\Etilde$ cluster were actually measured
in the appropriate laboratory basis, which differs from the
cluster basis when a single-qubit gate acts on the considered
qubit; referring to Eq. \eqref{Eexp}, this corresponds to the case
of qubits $4$, $5$ and $6$.
\begin{table}[t]
\caption{\label{table:results}Experimental probabilities of the obtained output states
for balanced and constant function, with (FF) or without (no-FF) feed-forward.
We indicate in bold character the data corresponding to
the expected outputs.}
\begin{ruledtabular}
\begin{tabular}[c]{c|rr|rr|}
&\multicolumn{2}{c}{$f_B$: Balanced}&\multicolumn{2}{c}{$f_C:$ Constant}\\
\hline
Output & No-FF$(\%)$ & FF$(\%)$& No-FF$(\%)$ & FF$(\%)$\\
\hline
$\ket{000}$ & $0.8\pm0.1$  & $0.9\pm0.1$  & $0.7\pm0.1$  & $0.9\pm0.1$  \\
$\ket{001}$ & $2.7\pm0.2$  & $2.6\pm0.1$  & ${\bf77.5\pm0.5}$ & ${\bf75.2\pm0.2}$ \\
$\ket{010}$ & $1.4\pm0.2$  & $1.4\pm0.2$  & $1.2\pm0.1$  & $1.4\pm0.1$  \\
$\ket{011}$ & $15.5\pm0.5$ & $14.1\pm0.1$ & $3.5\pm0.2$  & $3.4\pm0.1$  \\
$\ket{100}$ & $1.3\pm0.2$  & $1.0\pm0.1$  & $1.2\pm0.1$ & $1.0\pm0.1$  \\
$\ket{101}$ & $2.4\pm0.2$  & $3.4\pm0.1$  & $13.5\pm0.4$ & $14.1\pm0.2$ \\
$\ket{110}$ & $0.4\pm0.1$  & $1.4\pm0.1$  & $0.4\pm0.1$  & $1.4\pm0.1$  \\
$\ket{111}$ & ${\bf 75.5\pm0.6}$ & ${\bf75.2\pm0.2}$ & $2.0\pm0.2$  & $2.6\pm0.1$  \\
\end{tabular}
\end{ruledtabular}
\end{table}

The experimental results are in good agreement with
the theoretical predictions for both functions.
The main discrepancy resides on the output probabilities of the states
$\ket{011}$ for $f_B$ and $\ket{101}$ for $f_C$. These states differ from the expected outputs
in the value of the logical qubit $\bar1$.
{This is mainly due to the non perfect interference
visibility associated to the E/I momentum DOF ($V\sim 70\%$).
We attribute this to the difficulties in obtaining a perfect mode matching
in the second interferometer (see \cite{vall09qph} for more details).}

{\em Conclusions. --} We have presented an all-optical
implementation of the DJ algorithm for $n = 2$ qubits.
For this purpose, by taking advantage of the
generation of a six-qubit two-photon hyperentangled state, we
created a novel, high fidelity, two-dimensional six-qubit cluster
state that represents the first step for the realization of the
algorithm as a one-way quantum computation. We were then able to
evaluate a two-bit balanced function as well as a constant one and
to discriminate between them in one single run of the executed
program, in contrast to the three runs needed with
a classical computer. The correct output is identified at a
frequency of almost 1kHz without feed-forward,
a result which overcomes by several orders of magnitude what can
be achieved with a six-photon cluster state, according to the
current optical technology. By using all possible detection outputs and applying the
feed-forward corrections we could obtain a frequency 8 times
larger. Note that our experiment was actually performed with four
detectors \cite{vall09qph}. In order to consider all the possible
outcomes at the same time we would need 16 detectors.

The experimental results demonstrate the correctness of the proposed algorithm implementation
and represent the first proof of such a computation with a two-bit function
in the framework of the one-way model.


\acknowledgements We thank R. Jozsa for
useful discussions.




\end{document}